\documentclass{article}
\usepackage{spconf,amsmath,graphicx}
\usepackage{amssymb}
\usepackage{booktabs}
\usepackage{multirow}
\usepackage{cite}
\usepackage{setspace}
\usepackage{color}
\usepackage{makecell}
\title{Low algorithmic delay implementation of convolutional beamformer for online joint source separation and dereverberation}
\name{Kaien Mo$^1$, Xianrui Wang$^{1, 2}$, Yichen Yang$^{1, 2}$, Shoji Makino$^1$, and Jingdong Chen$^2$
\thanks{This work was supported by JSPS KAKENHI Grant Number 23H03423. }
}
\address{
$^1$Graduate School of Information, Production and Systems, \\Waseda University, Kitakyushu, Japan\\
$^2$Center of Intelligent Acoustics and Immersive Communications, \\
Northwestern Polytechnical University, Xi'an, China
}

\begin{document}
\ninept
\maketitle
\linespread{0.6}
\begin{abstract}
Blind-audio-source-separation (BASS) techniques, particularly those with 
low latency, play an important role in a wide range of real-time systems, 
e.g., hearing aids, in-car hand-free voice communication, real-time 
human-machine interaction, etc. Most existing BASS algorithms are deduced 
to run on batch mode, and therefore large latency is unavoidable. Recently, 
some online algorithms were developed, which achieve separation on a 
frame-by-frame basis in the short-time-Fourier-transform (STFT) domain and 
the latency is significantly reduced as compared to those batch methods. 
However, the latency with these algorithms may still be too long for many 
real-time systems to bear. To further reduce latency while achieving good 
separation performance, we propose in this work to integrate a weighted 
prediction error~(WPE) module into a non-causal sample-truncating-based 
independent vector analysis (NST-IVA). The resulting algorithm can maintain 
the algorithmic delay as NST-IVA if the delay with WPE is 
appropriately controlled while achieving significantly better performance, 
which is validated by simulations.   
\end{abstract}


\section{Introduction}

Blind audio source separation (BASS) refers to the problem of separating 
audio source signals from observed mixtures with minimal prior 
information\cite{makino2018audio, benesty2018fundamentals, 
huang2022fundamental, Wang2023MIBO}. Many methods have been developed to 
tackle this problem, among which the so-called independent vector 
analysis~(IVA) \cite{kim2006independent, hiroe2006IVA} has been widely 
investigated and has demonstrated promising separation performance. 
Originally, IVA-based algorithms are deduced to run on batch mode, and 
therefore large latency is unavoidable, which is unacceptable in most real 
applications. To achieve low latency, online versions of IVA are developed 
\cite{kim2010real, taniguchi2014auxiliary, 
nakashima2022inverse,ono2011stable}. This type of algorithms achieves 
separation on a frame-by-frame basis in the short-time-Fourier-transform 
(STFT) domain and the latency is significantly reduced as compared to those 
batch methods. However, the delay introduced by these algorithms may still be 
too long for many real-time applications since it depends on the frame length. 
The frame length in these algorithms has to be longer than the room 
impulse responses to achieve reasonably good separation performance. 

To address this issue, a group of methods called convolutional beamformer~(CBF) \cite{nakatani2020jointly, nakatani2020computationally, Yang2023geometrically}, which combine IVA and weighted prediction error~(WPE) techniques \cite{nakatani2010speech, yoshioka2012generalization}, is extended to the online version \cite{ueda2021low_IVA, ueda2021low_IVE, ueda2023constant}. By integrating WPE, the CBF methods can process both current and past frames to 
mitigate the impact of reverberation even when the STFT frame length is short 
\cite{ueda2021low_IVA}. A drawback of this type of algorithms is that they 
are computationally expensive as the convolutional operation uses a number of 
frames to achieve dereverberation. In this work, we introduce a method 
that combines online CBF \cite{ueda2021low_IVA} and non-causal sample-truncating-based 
independent vector analysis (NST-IVA) 
\cite{sunohara2017low}. This method uses a long STFT analysis window to 
reduce the length of convolutional operation in updating dereverberation filters while 
maintaining low algorithmic delay by truncating the non-causal samples of the 
process filters.  Simulation results show that the proposed system is able to 
reduce the algorithmic delay to as low as $4$ ms while producing better 
separation performance than its conventional counterparts. 

\section{Non-causal sample-truncating-based IVA}

\subsection{Algorithmic delay description}

\begin{figure}[t!]
\centering
\includegraphics[width = .48\textwidth]{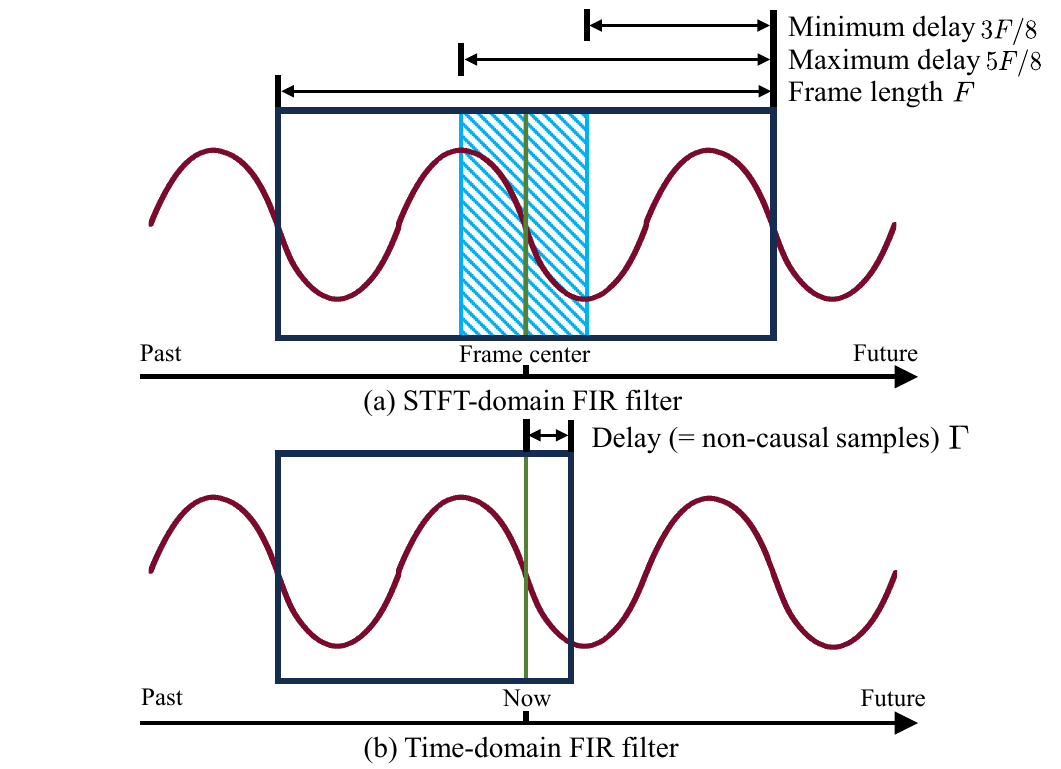}
\caption{Illustration of algorithmic delay of the STFT- (with window shift size of $F/4$) and time-domain methods.}
\label{latency_definition}
\end{figure}

The algorithmic delay of the STFT- and time-domain algorithms is illustrated in
Fig.~\ref{latency_definition}. In the STFT-domain, due to the use of 
overlap-add technique, the output at time $t$ is affected by multiple frames. 
To clarify the discussion, we calculate the algorithmic delay of an output point from the frame with the closest center to it. 
For example, as illustrated in Fig.~\ref{latency_definition}(a), if we assume that the 
frame size is $F$ and the overlap rate is $75\%$, the algorithmic delay of 
the samples within the light blue box depends on the frame noted by the black 
box. { The average delay of these samples is F/2 since STFT cannot be performed until all samples in this frame are received. In other words, the algorithmic delay of an STFT-domain filter is related to the analysis window length. In comparison, for the time-domain algorithms, as illustrated in Fig.~1(b), the algorithmic delay of the filter depends only on its non-causal sample number $\Gamma$. If truncating the non-causal components of a time-domain filter is feasible, its algorithmic delay can be reduced. }

\subsection{Conventional online IVA algorithms}

Suppose that there are $N$ sources in the sound field and we use an array of 
$M$ microphones to pick up the signals. The observation signals in the STFT 
domain can be expressed as 
\begin{equation}
    \label{FD_MTF_mixing_eq}
    \mathbf{x}(i,f) = \mathbf{A}(i;f)\mathbf{s}(i,f),
\end{equation}
where 
\begin{align}
\label{sig_vec_def}
\mathbf{x}(i,f) &= [{X}_{1}(i,f)\ \ {X}_{2}(i,f)\ \ \cdots\ \ {X}_{M}(i,f)]^\mathsf{T} \in \mathbb{C}^{M\times 1},\\
\mathbf{s}(i,f) &= [{S}_{1}(i,f)\ \ {S}_{2}(i,f)\ \ \cdots\ \ {S}_{N}(i,f)]^\mathsf{T} \in \mathbb{C}^{N\times 1},
\end{align}
are the observed and source signal vectors respectively, $f$ is the STFT bin 
index, $i$ is the time frame index, $\mathbf{A}(i;f)\in\mathbb{C}^{M\times 
N}$ is the mixing matrix, and $(\cdot)^{\mathsf{T}}$ denotes the transpose 
operation. In this paper, we only consider the case with two microphones 
and two sources, i.e., $N = M = 2$. In our future work, we will investigate the causality of the separation filter with more input channels and apply this method to a larger microphone array.
Assume that the $\mathbf{A}(i;f)$ matrix is of full 
rank, the source signals can be estimated with a linear filter, i.e.,  
\begin{equation}
    \label{FD_MTF_filter_def}
    \mathbf{y}(i,f)=\mathbf{W}(i;f)\mathbf{x}(i,f),
\end{equation}
where $\mathbf{W}(i;f)=\mathbf{A}^{-1}(i;f)$ is the separation matrix,  
$\mathbf{y}(i,f) = [{Y}_{1}(i,f)\ \ {Y}_{2}(i,f)\ \ \cdots\ \ 
{Y}_{N}(i,f)]^\mathsf{T} \in \mathbb{C}^{N\times 1}$ is an estimate of the 
source signals vector. 

The algorithmic delay in the STFT domain is bounded to the STFT frame length. 
One way to reduce the delay is through truncation \cite{sunohara2017low}. By 
inverse discrete Fourier transform (IDFT), the separation matrix 
$\mathbf{W}(i;f)$ is converted back to the time domain as 
\begin{equation}
    \label{IDFT_w}
    {w}_{n,m}(i;\tau)=\frac{1}{F}\sum_{f=-F/2}^{F/2-1}{W}_{n,m}(i;f)e^{j2\pi f\tau/F},
\end{equation}
where ${W}_{n,m}(i;f)$ is the $(n, m)$th element of $\mathbf{W}(i;f)$, and 
${w}_{n,m}(i;\tau)$ corresponds to the time-domain FIR filter coefficient of $n$th source, $m$th input channel with the discrete-time index $\tau 
\in [-F/2,F/2-1]$. Then, the separation process is expressed as 
\begin{equation}
    \label{TD_demix_def}
    {y}_{n}(t) = \sum_{\tau=-F/2}^{F/2-1}\sum_{m=1}^{M}{w}_{n,m}(i;\tau){x}_{m}(t-\tau),
\end{equation}
where $y_{n}(t)$ is the time-domain estimated signal of $n$th source at time 
$t$, $x_{m}(t)$ denotes the signal observed  at the $m$th microphone. As 
proved in \cite{sunohara2017low}, the ideal separation filter $w_{n,m}(i;\tau)$ 
is a causal filter when $N = M = 2$. Hence, the algorithm delay can be 
reduced through truncation. Suppose that a total of $\Gamma$ non-causal 
samples are truncated, the separated signal is expressed as   
\begin{equation}
    \label{TD_demix_def}
    {y}_{n}(t) = \sum_{\tau=-F/2+\Gamma}^{F/2-1}\sum_{m=1}^{M}{w}_{n,m}(i;\tau){x}_{m}(t-\tau).
\end{equation}
The algorithm delay is then shortened from $F/2$ to $F/2-\Gamma$.

\begin{figure}[t!]
\centering
\includegraphics[width = .48\textwidth]{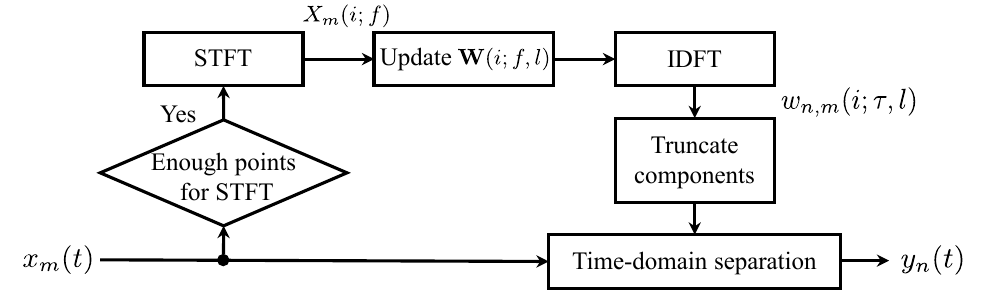}
\caption{Flowchart of the proposed algorithm.}
\label{struct_flow}
\end{figure}

\section{Proposed method}

\subsection{System structure}

The flowchart of the proposed algorithm is shown in Fig.~\ref{struct_flow}. 
To further improve the BASS performance in heavy reverberation while 
maintaining a low algorithmic delay, a parallel processing structure similar 
to the one in  NST-IVA \cite{sunohara2017low} is used. This structure updates 
the joint separation and dereverberation filters by the conventional online 
CBF in the STFT-domain \cite{ueda2021low_IVA}. After updating, the filters are transformed back to 
the time-domain to separate new observed signals directly. 

\subsection{Filters updating in the STFT domain}

\subsubsection{Problem formulation and probabilistic model}

When the room impulse response is much longer than the STFT frame length, the 
instantaneous mixing model is not sufficient. Therefore, we adopt the 
convolutional transfer function (CTF) model in CBF \cite{nakatani2020jointly, 
nakatani2020computationally} to model the problem, i.e., 
\begin{equation}
    \label{FD_mixing_eq}
    \mathbf{x}(i,f) = \sum_{l_{A}=0}^{L_A-1}\mathbf{A}(i;f,l_{A})\mathbf{s}(i-l_{A},f),
\end{equation}
where $\mathbf{A}(i;f,l_{A})\in\mathbb{C}^{M\times N}$ is the convolutional 
mixing matrix at time lag $l_{A}$, and $L_A$ is the order of the mixing 
filters. Correspondingly, the convolutional filters are used for simultaneous 
dereverberation and separation, i.e., 
\begin{equation}
    \label{filter_def}
    \mathbf{y}(i,f)=\mathbf{W}(i;f,0)\mathbf{x}(i,f)+\sum^{L+D-1}_{l=D}\mathbf{W}(i;f,l)\mathbf{x}(i-l,f),
\end{equation}
where $\mathbf{W}(i;f,l)\in\mathbb{C}^{N\times M}$ is a filter with time lag 
$l$, $L$ is the total orders of filters, and $D$ is a delay parameter 
introduced to prevent from distortion \cite{nakatani2010speech}. Based on the 
online source-wise factorization of CBF \cite{ueda2021low_IVA}, the filters 
in \eqref{filter_def} can be decomposed into the following two processes: 
\begin{align}
    \label{sw_wpe_def}
    \mathbf{z}_{n}(i,f) &= \mathbf{x}(i,f) - \mathbf{G}^{\mathsf{H}}_{n}(i;f)\overline{\mathbf{x}}(i,f),\\
    \label{sw_ss_def}
    {Y}_{n}(i,f) &= \mathbf{q}^\mathsf{H}_{n}(i;f)\mathbf{z}_{n}(i,f).
\end{align}
The first process in (\ref{sw_wpe_def}) corresponds to the dereverberation 
process of the $n$th source, where $\mathbf{G}_{n}(i;f)$ and 
$\mathbf{z}_{n}(i,f)$ are the dereverberation filter and the output signal, 
$(\cdot)^{\mathsf{H}}$ represents conjugate transpose and 
$\overline{\mathbf{x}}(i,f)=[\mathbf{x}^{\mathsf{T}}(i-D,f)\ \ 
\mathbf{x}^{\mathsf{T}}(i-D-1,f)\ \ \cdots\ \ 
\mathbf{x}^{\mathsf{T}}(i-D-L+1,f)]^{\mathsf{T}}$ is the vector containing 
past samples of the mixture signals. The second process in \eqref{sw_ss_def} corresponds to the 
separation process, which uses filter $\mathbf{q}_{n}(i;f)$ to extract the 
$n$th source signal. Substituting \eqref{sw_wpe_def} and \eqref{sw_ss_def} 
into \eqref{filter_def}, we obtain the relation between $\mathbf{W}(i;f,l)$ 
and $\mathbf{G}_{n}(i;f)$, $\mathbf{q}_{n}(i;f)$ , i.e.,  
$\mathbf{w}_{n}(i;f,0)=\mathbf{q}_{n}(i;f)$ and 
$[\mathbf{w}^{\mathsf{T}}_{n}(i;f,D),\cdots,\mathbf{w}^{\mathsf{T}}_{n}(i;f,L+D-1)]^{\mathsf{T}}=-\mathbf{G}_{n}(i;f)\mathbf{q}_{n}(i;f)$, 
where $\mathbf{w}_{n}(i;f,l)$ is the $n$th column of 
$\mathbf{W}^{\mathsf{H}}(i;f,l)$. Note that even though an STFT-domain 
separated signal component $Y_{n}(i,f)$ is produced in the updating process 
of $\mathbf{W}(i;f,l)$, this component is not be transformed to the time-domain as the final separated signal. 

{ To deal with the joint dereverberation and source separation problem,} the CBF method assumes that the source 
signal in the STFT-domain follows a Gaussian distribution with zero mean and 
time-dependent variance $\sigma_{n}(i)=\mathbb{E}(|{S}_{n}(i,f)|^2)$, i.e.,  
\begin{equation}
    \label{Gaussain}
    {S}_{n}(i,f)\sim\mathcal{N}(0,\sigma_{n}(i)).
\end{equation}
To estimate the adaptive filters to count for time-variant effects, the 
online version of CBF \cite{ueda2021low_IVA} adds a forgetting factor $\beta$ 
to the conventional CBF's negative log-likelihood function as 
\begin{align}
\label{o_CBF_cost}
\mathcal{L}(\mathbf{\Theta}(i))=&-2\sum_{f}\log|\det \mathbf{Q}(i;f)|\nonumber\\
&+\frac{\sum_{n,f,i'\leq i}{\beta}^{i-i'}\left( \log \sigma_{n}(i')\!+\!\frac{|{Y}_{n}(i',f)|^2}{\sigma_{n}(i')} \right)}{\sum_{i'\leq i}{\beta}^{i-i'}},
\end{align}
where $\mathbf{Q}(i;f)=[\mathbf{q}_{1}(i;f)\ \ \mathbf{q}_{2}(i;f)\ \ \cdots\ \ \mathbf{q}_{N}(i;f)]^{\mathsf{H}}\in\mathbb{C}^{N\times M}$ is the separation matrix, and $\Theta(i)=\{\Theta_{\sigma}(i),\Theta_{\mathbf{G}}(i),\Theta_{\mathbf{Q}}(i)\}$ is the unknown parameter set, $\Theta_{\sigma}(i) = \{\sigma_{n}(i)\}$, $\Theta_{\mathbf{G}}(i)=\{\mathbf{G}_{n}(i;f)\}$, and $\Theta_{\mathbf{Q}}(i)=\{\mathbf{Q}_{n}(i;f)\}$. Each parameter set can be updated iteratively using the coordinate ascent method \cite{yoshioka2010blind}. 
\vspace{-0.30cm}
\subsubsection{Update of $\Theta_{\sigma}(i)$}

The process of filter update begins with calculating the estimated source 
signal $\mathbf{y}(i,f)$ using \eqref{sw_wpe_def} and \eqref{sw_ss_def} with the filters $\mathbf{G}_{n}(i-1;f)$ and 
$\mathbf{Q}(i-1;f)$ being updated in the previous frame. Then, online CBF estimates the variance as 
\begin{equation}
    \sigma_{n}(i)\leftarrow\sum_{f=-F/2}^{F/2-1}|{Y}_{n}(i,f)|^{2}/F.
\end{equation}
\vspace{-0.30cm}
\subsubsection{Update of $\Theta_{\mathbf{G}}(i)$}

Following the method in \cite{ueda2021low_IVA}, one can update 
$\Theta_{\mathbf{G}}(i)$ through minimizing \eqref{o_CBF_cost} while fixing 
$\Theta_{\sigma}(i)$ and $\Theta_{\mathbf{Q}}(i)$, i.e,  
\begin{equation}
    \label{G_update}
    \mathbf{G}_{n}(i;f)\leftarrow\mathbf{R}^{-1}_{n}(i;f)\mathbf{P}_{n}(i;f),
\end{equation}
where 
\begin{align}
\label{R_update}
    \mathbf{R}_{n}(i;f) &= \beta\mathbf{R}_{n}(i-1;f)+\frac{\overline{\mathbf{x}}(i,f)\overline{\mathbf{x}}^{\mathsf{H}}(i,f)}{\sigma_{n}(i)},\\
    \label{P_update}
    \mathbf{P}_{n}(i;f) &= \beta\mathbf{P}_{n}(i-1;f)+\frac{\overline{\mathbf{x}}(i,f){\mathbf{x}}^{\mathsf{H}}(i,f)}{\sigma_{n}(i)},
\end{align}
are two spatio-temporal covariance matrices.

To achieve real-time processing, the matrix inversion lemma \cite{diniz1997adaptive} is applied to promote the computational efficiency of \eqref{G_update}. Hence, the calculation of $\mathbf{R}_{i,f,n}^{-1}$ can be written as
\begin{align}
    \label{k_update}
    \hspace{-0.8mm} \mathbf{k}_{n}(i;f)\hspace{-0.8mm}&\leftarrow\hspace{-0.8mm}\frac{\mathbf{R}^{-1}_{n}\hspace{-0.5mm}(i\hspace{-0.8mm}-\hspace{-0.8mm}1;f)\overline{\mathbf{x}}(i,f)}{\beta \sigma_{n}(i)+\overline{\mathbf{x}}^\mathsf{H}(i,f)\mathbf{R}^{-1}_{n}\hspace{-0.5mm}(i\hspace{-0.8mm}-\hspace{-0.8mm}1;f)\overline{\mathbf{x}}(i,f)},\\
    \label{R_inv}
    \hspace{-0.8mm} \mathbf{R}^{-1}_{n}\hspace{-0.5mm}(i;f)\hspace{-0.8mm}&\leftarrow\hspace{-0.8mm}\frac{\mathbf{R}^{-1}_{n}\hspace{-0.5mm}(i\hspace{-0.8mm}-\hspace{-0.8mm}1;f)\hspace{-0.8mm}-\hspace{-0.8mm}\mathbf{k}_{n}(i;f)\overline{\mathbf{x}}^\mathsf{H}(i,f)\mathbf{R}^{-1}_{n}\hspace{-0.5mm}(i\hspace{-0.8mm}-\hspace{-0.8mm}1;f)}{\beta},
\end{align}
where $\mathbf{k}_{n}(i;f)$ is the Kalman gain. Substituting \eqref{P_update} 
and \eqref{R_inv} into \eqref{G_update} gives the following online update rule 
of $\mathbf{G}_{n}(i;f)$: 
\begin{equation}
    \mathbf{G}_{n}(i;f)\leftarrow \mathbf{G}_{n}(i-1;f)+\mathbf{k}_{n}(i;f)\mathbf{z}^{\mathsf{H}}_{n}(i,f).
\end{equation}
\vspace{-0.30cm}
\subsubsection{Update of $\Theta_{\mathbf{Q}}(i)$ }

If fixing other parameters, the cost function in \eqref{o_CBF_cost} 
degenerates to the one used in the online AuxIVA. Hence, the separation 
matrix $\mathbf{Q}_{n}(i;f)$ can be estimated through the iterative source 
steering (ISS)-based updating rules \cite{scheibler2020fast}, which has been 
used in many IVA-based methods \cite{nakashima2022inverse, 
goto2022geometrically, mo2023joint}. After initializing the separation matrix 
of the current frame $\mathbf{Q}_{n}(i;f)=\mathbf{Q}_{n}(i-1;f)$, this matrix 
can be updated with an auxiliary vector $\mathbf{v}_{k}(i;f)$ as 
\begin{equation}
    \label{Q_update_ISS}
    \mathbf{Q}(i;f) = \mathbf{Q}(i;f)-\mathbf{v}_{k}(i;f)\mathbf{q}^{\mathsf{H}}_{k}(i;f).
\end{equation}
This update process is repeated for $k=1,\dots,N$. To continue the updating, 
$\mathbf{v}_{k}(i;f)$ needs to be calculated. By substituting 
\eqref{Q_update_ISS} into \eqref{o_CBF_cost} and fixing other parameters, the 
update rules of $\mathbf{v}_{k}(i;f)$ can be derived as 
\begin{align}
\label{update_vnn}
    {V}_{n,k}(i;f) =
\begin{cases}
1-(\mathbf{q}^{\mathsf{H}}_{n}(i;f)\mathbf{U}_{n}(i;f)\mathbf{q}_{n}(i;f))^{-1/2}, \hspace{-2.8mm}& {\text{if~}}n = k,\\
\frac{\mathbf{q}^{\mathsf{H}}_{n}(i;f)\mathbf{U}_{n}(i;f)\mathbf{q}_{k}(i;f)}{\mathbf{q}^{\mathsf{H}}_{k}(i;f)\mathbf{U}_{n}(i;f)\mathbf{q}_{k}(i;f)}, \hspace{-2.8mm}&  \text{else},
\end{cases}
\end{align}
where ${V}_{n,k}(i;f)$ is the $n$th element of $\mathbf{v}_{k}(i;f)$, 
\begin{equation}
    \label{U_update}
    \mathbf{U}_{n}(i;f)=\alpha\mathbf{U}_{n}(i-1;f)+(1-\alpha)\frac{\mathbf{z}_{n}(i,f)\mathbf{z}^{\mathsf{H}}_{n}(i,f)}{\sigma_{n}(i)},
\end{equation}
is the weighted covariance matrix of the signal after dereverberation updated in an autoregressive manner \cite{taniguchi2014auxiliary} with a forgetting factor $\alpha$. 
\vspace{-0.47cm}
\subsection{Time-domain implementation}

To reduce the algorithmic delay, we convert the original STFT-domain 
convolutional filters $\mathbf{W}(i;f,l)$ back to the time domain as in the 
conventional NST-IVA.  Then, once a new signal sample is accessible, the 
output signal is generated immediately without waiting for enough samples for 
the next STFT frame. The transformation of $\mathbf{W}(i;f,l)$ by IDFT can be 
expressed as 
\begin{equation}
    \label{IDFT_w}
    {w}_{n,m}(i;\tau,l)=\frac{1}{F}\sum_{f=-F/2}^{F/2-1}{W}_{n,m}(i;f,l)e^{j2\pi f\tau/F},
\end{equation}
where ${W}_{n,m}(i;f,l)$ is the $(n, m)$th element of $\mathbf{W}(i;f,l)$ and 
${w}_{n,m}(i;\tau,l)$ corresponds to the time-domain FIR filter parameter. 
Since STFT is a linear process, the STFT-domain joint separation and 
dereverberation process in \eqref{filter_def} is equal to the following 
time-domain processing: 
\begin{align}
\label{TD_filter}
{y}_{n}(t)=&\sum_{\tau=-F/2}^{F/2-1}\sum_{m=1}^{M} \Big[{w}_{n,m}(i;\tau,0){x}_{m}(t-\tau)\Big.\nonumber\\
    \Big.&+\sum^{L+D-1}_{l=D}{w}_{n,m}(i;\tau,l){x}_{m}(t-\tau-l\delta)\Big],  
\end{align}
where $\delta$ is the window shift length of the STFT in updating filters. Moreover, we consider the window function ${h}(\tau)$ of the STFT with the discrete time index $\tau$ and rewrite the time domain processing \eqref{TD_filter} as 
\begin{align}
        \label{TD_filter_win}
    {y}_{n}(t)=&\sum_{\tau=-F/2}^{F/2-1}\sum_{m=1}^{M}\Big[{h}(\tau){w}_{n,m}(i;\tau,0){x}_{m}(t-\tau)\Big.\nonumber\\ 
    &\Big. +\sum^{L+D-1}_{l=D}{h}(\tau){w}_{n,m}(i;\tau,l){x}_{m}(t-\tau-l\delta)\Big].
\end{align}
To reduce the algorithmic delay, the non-causal samples of the filter 
$w_{n,m}(i;\tau,l)$ need to be truncated. If $D\times\delta \ge F/2$, all 
non-causal samples fall only in the separation filter  ${w}_{n,m}(i;\tau,0)$; 
so, in this work, we consider only truncating ${w}_{n,m}(i;\tau,0)$. Besides, 
since the first tap of the convolutional transfer function 
$\mathbf{W}(i;f,0)$ is equal to the instant separation matrix in online IVA 
\cite{taniguchi2014auxiliary} as shown in subsection 3.2, the causality of 
${w}_{n,m}(i;\tau,0)$ can be proved in a same way as \cite{sunohara2017low}. 
Hence, the non-causal samples of ${w}_{n,m}(i;\tau,0)$ can be truncated 
without suffering heavy performance degradation. Suppose that a total of 
$\Gamma$ non-causal samples are truncated, the proposed FIR filter process 
\eqref{TD_filter_win} can be written as 
\begin{align}
        \label{TD_filter_trun}
    {y}_{n}(t)=&\sum_{m=1}^{M}\Bigl[\sum_{\tau=-F/2+\Gamma}^{F/2-1}{h}(\tau){w}_{n,m}(i;\tau,0){x}_{m}(t\hspace{-0.8mm}-\hspace{-0.8mm}\tau)\Bigl.\nonumber\\
    \Bigl.&+\sum_{\tau=-F/2}^{F/2-1}\sum^{L+D-1}_{l=D}{h}(\tau){w}_{n,m}(i;\tau,l){x}_{m}(t\hspace{-0.8mm}-\hspace{-0.8mm}\tau\hspace{-0.8mm}-\hspace{-0.8mm}l\delta)\Bigl].
\end{align}
With this step, the algorithmic delay is shortened to $F/2-\Gamma$ samples as 
in NST-IVA. 

\section{Experimental evaluation}

\subsection{Experiment setup}

The clean speech signals for simulations are taken from the ATR Japanese 
Speech Database \cite{kurematsu1990atr}. Each mixed signal is generated with 
two source signals arbitrarily selected from two speakers from the database. 
If the source signal is less than $20$ s, it is concatenated with another 
selected source signal and truncated then so the overall length is $20$ s. 
The \textit{pyroomacoustics} Python package \cite{scheibler2018pyroomacoustics} is used to simulate room impulse 
responses (RIRs) where the room boundary coefficients are controlled so that the 
room reverberation time, i.e., $T_{60}$, is $600$~ms. We simulate a 
two-element microphone array with a spacing of $2$~cm to observe the signals. 
The proposed method is validated under two different pairs of incidence 
angles in the two-speaker scenarios: ($30$°, $90$°) and ($30$°, $150$°). For 
each condition,  $12$ simulations with different mixed signals are performed 
and the average results are used for evaluation and comparison. The distance 
between the microphone array center and the sources is $2$ m. All signals are 
sampled at $16$ kHz. 

Two conventional methods are used as the baseline systems: NST-IVA 
\cite{sunohara2017low} (denoted as TD-IVA) and the STFT-domain online CBF 
\cite{ueda2021low_IVA} (denoted as FD-CBF). The proposed time-domain implementation of CBF is denoted as TD-CBF. Hann window is used as the analysis 
window in STFT and the overlap ratio is set to $75\%$. The frame length, the 
number of truncated samples, and the corresponding algorithmic delay of the 
studied methods are shown in Table~1. The values of $D$ and 
$L$ in CBF are set to $2$ and $10$ for the case with $64$-ms frame length 
while $8$ and $10$ for FD-CBF. Although setting a longer WPE filter length, 
i.e., a larger value of $L$, can help achieve better performance when using 
shorter STFT windows, this would increase the computation cost. Therefore, we 
only used the same WPE filter length for comparison. The forgetting factor of 
IVA $\alpha$ and WPE $\beta$ are set to $0.99$ and $0.999$ respectively. All 
simulations were conducted on a workstation powered by Intel Xeon E3-1505M. 

The improvement of source-to-distortion ratio ($\Delta$SDR), 
source-to-interferences ratio ($\Delta$SIR), and sources-to-artifacts ratio 
($\Delta$SAR)   \cite{vincent2006performance} are used as the metrics to 
evaluate the separation performance. 

\begin{table}[t!]\footnotesize
\begin{center}
\setlength{\abovecaptionskip}{0pt}
\setlength{\belowcaptionskip}{10pt}
\caption{The process setting and the algorithmic delay of studied methods}
\vspace{0.1cm}
\begin{tabular}{c|rrr}
\toprule
\multirow{2}{*}{Method} & \makecell[c]{Window} & \makecell[c]{Truncated} & \makecell[c]{Algorithmic} \\
         & \makecell[c]{length} & \makecell[c]{points} & \makecell[c]{delay} \\
\midrule
TD-IVA (32 ms)& $64$ ms & $0$ & $32$ ms\\
FD-CBF (4 ms)& $8$ ms & $0$ & $4$ ms\\
\textbf{TD-CBF (Proposed, 32 ms)}& $64$ ms & $0$ & $32$ ms\\
\textbf{TD-CBF (Proposed, 4 ms)}& $64$ ms & $448$ & $4$ ms\\
\bottomrule
\end{tabular}
\end{center}
\label{latency_table}
\end{table}

\subsection{Experiment result}

\begin{figure}[t!]
\centering
\includegraphics[width = .48\textwidth]{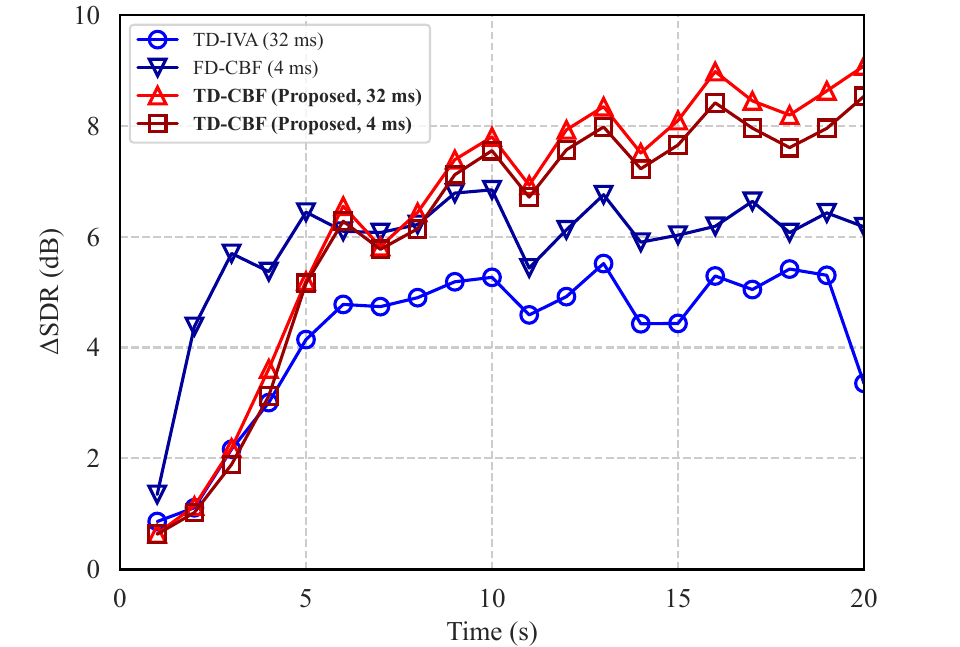}
\vspace{-0.20cm}
\caption{Separation performance vs time.}
\label{line-plot}
\end{figure}
\begin{figure}[t!]
\centering
\includegraphics[width = .48\textwidth]{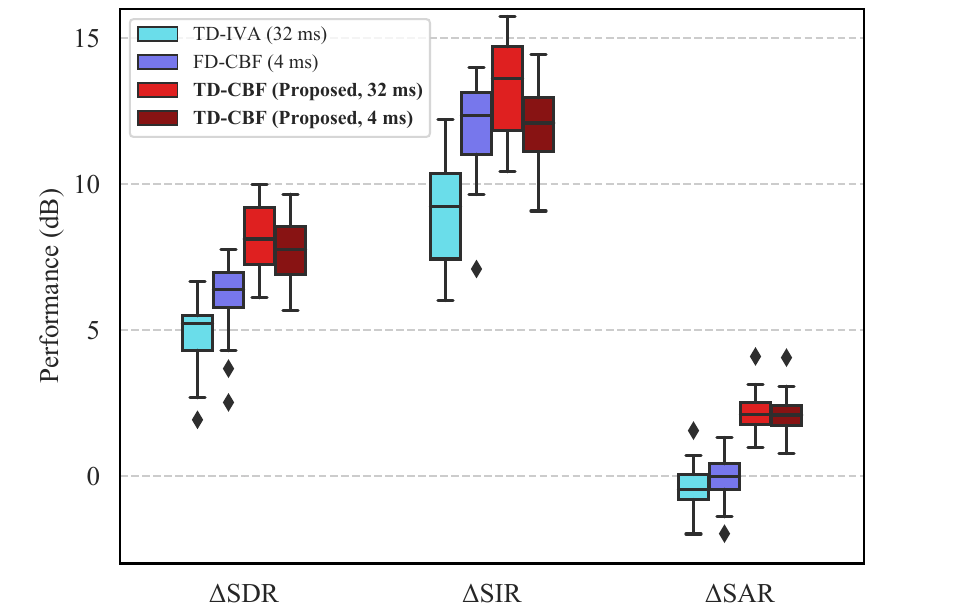}
\vspace{-0.20cm}
\caption{Separation performance after convergence.}
\label{box-plot}
\end{figure}

The $\Delta$SDR results as a function of time for all the studied methods are 
plotted in Fig.~\ref{line-plot}. As seen, all the studied methods converge at 
$10$ s. For a fair comparison of performance, we compute the average 
$\Delta$SDR, $\Delta$SIR, and $\Delta$SAR after all the methods converge, 
i.e., the results for the first $10$-s are discarded. The results are shown 
in Fig.~\ref{box-plot}. 

As seen in Fig.~\ref{line-plot}, FD-CBF with an $8$~ms analysis window takes 
only 3 seconds to converge while the other methods with a 64~ms analysis 
window take 6 seconds to converge. This shows that the methods converge 
slower with longer STFT window length than with shorter window length. 
According to this result, in practical systems, setting a relatively short 
window at the beginning and gradually increasing the length should be an 
appropriate way to achieve both good convergence speed and separation 
performance. From the results in Fig.~\ref{box-plot}, the proposed TD-CBF 
method (integrated with WPE dereverberation) yields much higher $\Delta$SDR, 
$\Delta$SIR, and $\Delta$SAR results than TD-IVA after convergence. In 
comparison with FD-CBF, the proposed method can maintain the same algorithmic 
delay ($4$ ms) by truncating 448 non-causal samples, while using a relatively 
long STFT window length, which makes it more effective to deal with heavy 
reverberation. As a result, the proposed method also demonstrates better 
$\Delta$SDR, $\Delta$SAR than FD-CBF with the same algorithmic delay. 

\section{Conclusion}
To achieve source recovery in reverberant environments with low latency, this 
paper developed an algorithm that combines the idea of non-causal sample 
truncation and WPE dereverberation method. Due to the application of 
non-causal sample truncation, the deduced algorithm is able to control the 
algorithmic delay as small as 4 ms. Meanwhile, due to the application of WPE, 
the algorithm is able to achieve better speech separation performance in strong 
reverberant environments in terms of source-to-distortion ratio and 
source-to-interferences ratio. 


\newpage 
\bibliographystyle{IEEEtran}

\end{document}